\documentclass{PoS}
\usepackage{amsmath}
\usepackage{amssymb}
\usepackage{graphicx}
\usepackage{cite}
\usepackage{wrapfig}
\usepackage{overpic}
\newcommand{\kt}{{\mathbf k}}
\newcommand{\qt}{{\mathbf q}} 

\newcommand{\ptt}{\mathbf{\tilde{p}}}
\newcommand{\qtt}{\mathbf{\tilde{q}}}
\newcommand{\Pbb}{\mathbb{P}}

\title{Pgg TMD splitting function in kT factorization}
\ShortTitle{Pgg TMD splitting function in kT factorization}

\author{\speaker{A.~Kusina}\\ 
        The H. Niewodnicza\'nski Institute of Nuclear Physics, Polish Academy of Sciences, \\
        ul. Radzikowskiego 152, 31-342, Cracow, Poland\\
        E-mail: \email{Aleksander.Kusina@ifj.edu.pl}}
\author{M.~Hentschinski\\
       Departamento de Actuaria, F\'isica y Matem\'aticas,
       Universidad de las Americas Puebla, \\ Santa Catarina Martir, 72820
       Puebla, Mexico}
\author{K.~Kutak\\
       The H. Niewodnicza\'nski Institute of Nuclear Physics, Polish Academy of Sciences, \\
       ul. Radzikowskiego 152, 31-342, Cracow, Poland}
\author{M.~Serino\\ 
Department of Physics,  Ben Gurion University of the Negev,\\ Beer Sheva 8410501, Israel}

\abstract{We calculate the transverse momentum dependent gluon-to-gluon splitting function
within $k_T$-factorization~\cite{Hentschinski:2017ayz},
generalizing the framework of~\cite{Curci:1980uw,Catani:1994sq}
and extending our previous works~\cite{Gituliar:2015agu, Hentschinski:2016wya} for the quark
splittings. While existing versions of $k_T$ factorized evolution equations contain already
a gluon-to-gluon splitting function i.e. the leading order BFKL kernel or the CCFM kernel,
the obtained splitting function has the important property that it reduces to the leading
order BFKL kernel in the low $z$ limit, to the DGLAP gluon-to-gluon splitting function
in the collinear limit as well as to the CCFM kernel in the angular ordered region.}

\FullConference{XXVI International Workshop on Deep-Inelastic Scattering and Related Subjects (DIS2018)\\
		16-20 April 2018\\
		Kobe, Japan}

\begin{document}

\section{Introduction}
Parton distributions functions (PDFs) are crucial elements of collider
phenomenology. In presence of a hard scale $M$ with
$M \gg \Lambda_{\text{QCD}}$,
factorization theorems allow to express cross-sections as convolutions of parton
densities (PDFs) and hard matrix elements, where the latter are calculated
within perturbative QCD~\cite{Collins:1984xc}.  This was first
achieved within the framework of collinear
factorization~\cite{Ellis:1978ty,Collins:1985ue,Bodwin:1984hc}, where the
incoming partons are taken to be collinear with the respective mother
hadron.  Calculating hard matrix elements to higher orders in the strong
coupling constant,  one can systematically improve the precision of the
theoretical prediction, by incorporating more loops and more
emissions of real partons.  These extra emissions allow to
improve the kinematic approximation inherent to the leading order (LO)
description.
As an alternative to improving the kinematic description through the
calculation of higher order corrections, one may attempt
to account for the bulk of kinematic effects already at leading order. 
An important example of such kinematic effects is the transverse momentum
$k_T$ of the initial state partons, which is set to zero within collinear
factorization.
Schemes which
provide an improved kinematic description already at the leading order
involve in general Transverse-Momentum-Dependent (TMD) or
`unintegrated' PDFs~\cite{Angeles-Martinez:2015sea}.
TMD PDFs arise naturally in regions of phase space characterized by a hierarchy
of scales. A particularly interesting example is provided by the so called low $x$
region, where $x$ is the ratio of the hard scale $M^2$ of the process and the
center-of-mass energy squared $s$. The low $x$ region corresponds therefore
to the hierarchy $s \gg M^2 \gg \Lambda_{\text{QCD}}^2$. In such a kinematical
setup, large logarithms $\ln 1/x$ can compensate for the smallness of the
perturbative strong coupling $\alpha_s $ and it is necessary to resum
terms $\left(\alpha_s \ln 1/x \right)^n$ to all orders to maintain the predictive
power of the perturbative expansion. Such a resummation is achieved by the
Balitsky-Fadin-Kuraev-Lipatov (BFKL)~\cite{Fadin:1975cb,Kuraev:1976ge,Kuraev:1977fs,Balitsky:1978ic}
evolution equation.
Its formulation is based on the so called $k_T$ (or high-energy)
factorization~\cite{Catani:1990eg} which is strictly speaking valid in the
high energy limit, $s \gg M^2$. In this approach one obtains QCD cross-sections as
convolutions (in transverse momentum) of unintegrated gluon density and $k_T$-dependent
perturbative coefficients.


While high energy factorization provides a well defined calculational framework
the applicability of the results is naturally limited to the low $x$
limit.
If the ensuing formalism is naively extrapolated to intermediate or large
$x$, the framework is naturally confronted with a series of
problems and short-comings, e.g. contributions of quarks to
the evolution arise as a pure next-to-leading order (NLO) effect and
elementary vertices violate energy conservation {\it i.e.}
conservation of the longitudinal momentum fraction.
One can account for such effects by including a resummation of terms
which restore subleading, but numerically relevant, pieces of the
Dokshitzer-Gribov-Lipatov-Altarelli-Parisi (DGLAP)
\cite{Gribov:1972ri,Altarelli:1977zs,Dokshitzer:1977sg} splitting functions
\cite{Salam:1998tj,Altarelli:2005ni,Ciafaloni:2003kd,SabioVera2005,Hentschinski:2012kr,Hentschinski:2013id,Kwiecinski:1997ee,Bonvini:2017ogt}.
Even though these resummations have
been successful in stabilizing low $x$ evolution into the region of
intermediate $x \sim 10^{-2}$, extrapolations to larger values of $x$ are still prohibited. 
Moreover, by merely resumming and calculating higher order corrections within the BFKL formalism, 
one essentially repeats the program initially outlined for collinear factorization:
higher order corrections are calculated to account for kinematic effects which are beyond the regarding factorization scheme.
To arrive at a framework which avoids the need to account for
kinematic effects through the calculation of higher order corrections,
it is therefore necessary to devise a scheme which accounts for
both DGLAP (conservation of longitudinal momentum) and
BFKL (conservation of transverse momentum) kinematics. Note that
the mere definition of such a scheme is difficult: neither the hard
scale of the process (as in DGLAP evolution) nor $x$ (BFKL evolution)
provides at first a suitable expansion parameter, if one
desires to keep exact kinematics in both variables. To overcome these
difficulties, we follow here a proposal initially outlined
in~\cite{Catani:1994sq}. There, the low $x$ resummed DGLAP
splitting functions have been constructed following the definition of
DGLAP splittings by Curci-Furmanski-Petronzio (CFP)~\cite{Curci:1980uw}.
The authors of~\cite{Catani:1994sq} were able to define a TMD
gluon-to-quark splitting function $\tilde{P}_{qg}$, both
exact in transverse momentum and longitudinal momentum fraction.%
    \footnote{Hereafter, we will use the symbol $\tilde{P}$ to indicate a
    transverse momentum dependent splitting function.}
Following observation of~\cite{Hautmann:2012sh} two of us generalized this scheme to
calculate the remaining splittings which involve quarks, $\tilde{P}_{gq}$,
$\tilde{P}_{qg}$ and $\tilde{P}_{qq}$~\cite{Gituliar:2015agu}. 
The computation of the gluon-to-gluon splitting $\tilde{P}_{gg}$ required
a further modification of the formalism used in~\cite{Catani:1994sq,Gituliar:2015agu}
which we did recently in~\cite{Hentschinski:2017ayz}.
In this contribution we summarize the most relevant results obtained in this work.

\section{Calculation of gluon-to-gluon splitting}
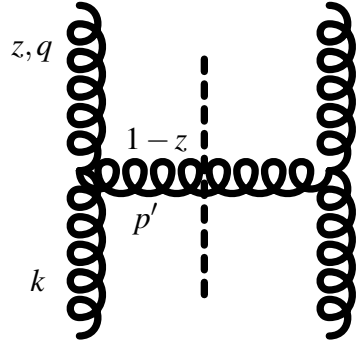
\begin{wrapfigure}{r}{0.4\textwidth}
\centering
\vspace{-0.5cm}
\begin{overpic}[width=0.29\textwidth]{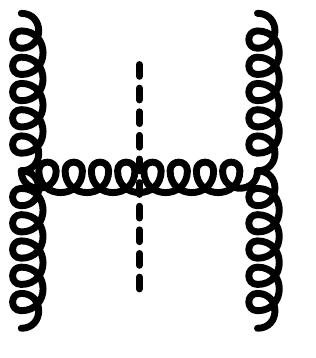}
 \put(-7,20){{\large $k$}}
 \put(-12,83){{\large $z,q$}}
 \put(20,37){{\large $p'$}}
 \put(18,58){{\large $1-z$}}
\end{overpic}
\vspace{-0.5cm}
\caption{Diagram contributing to the real $\tilde{P}_{gg}$ splitting function at leading order.
}
\label{fig:pgg}
\end{wrapfigure}
The calculation of the $\tilde{P}_{gg}$ splitting function follows the prescription
used in our papers~\cite{Gituliar:2015agu,Hentschinski:2017ayz}.
We are working in the
high energy kinematics where the momentum of the incoming parton is off-shell
(see Fig.~\ref{fig:pgg}) and given by: $k^\mu = y p^\mu + k_\perp^\mu$, 
where $p$ is a light-like momentum defining the
direction of a hadron beam and $n$ defines the axial gauge with $n^2=p^2=0$
(that is necessary when using CFP inspired formalism).
Additionally, the outgoing momenta is parametrized as:
$q^\mu = x p^\mu + q_\perp^\mu + \frac{q^2+\qt^2}{2x p\cdot n} n^\mu$ and we also use:
$\qtt = \qt - z \kt$ with $z=x/y$.

The TMD splitting function, $\tilde{P}_{gg}$, is defined as 
\begin{equation}
\begin{split}
\label{eq:TMDkernelDefINI}
\hat K_{gg} \left(z, \frac{\kt^2}{\mu^2}, \epsilon \right) &=
z \int \frac{d^{2 + 2 \epsilon} {\qt}}{2(2\pi)^{4+2\epsilon}}
     \underbrace{\int d q^2 \, \mathbb{P}_{g,\,\text{in}} \otimes
                 \hat{K}_{gg}^{(0)}(q, k) \otimes \mathbb{P}_{g,\,\text{out}}}
                    _{\tilde{P}_{gg}^{(0)} \left(z, \kt, \qtt, \epsilon \right)}
     \,\Theta(\mu_F^2+q^2),
\end{split}
\end{equation}
with $\mathbb{P}_{g}$ being appropriate projection operators and $\hat{K}_{gg}$
the matrix element contributing to the kernel, which at LO is given by the diagram
of Fig.~\ref{fig:pgg}.

\begin{figure}[ht]
\begin{center}
\includegraphics[scale=0.8]{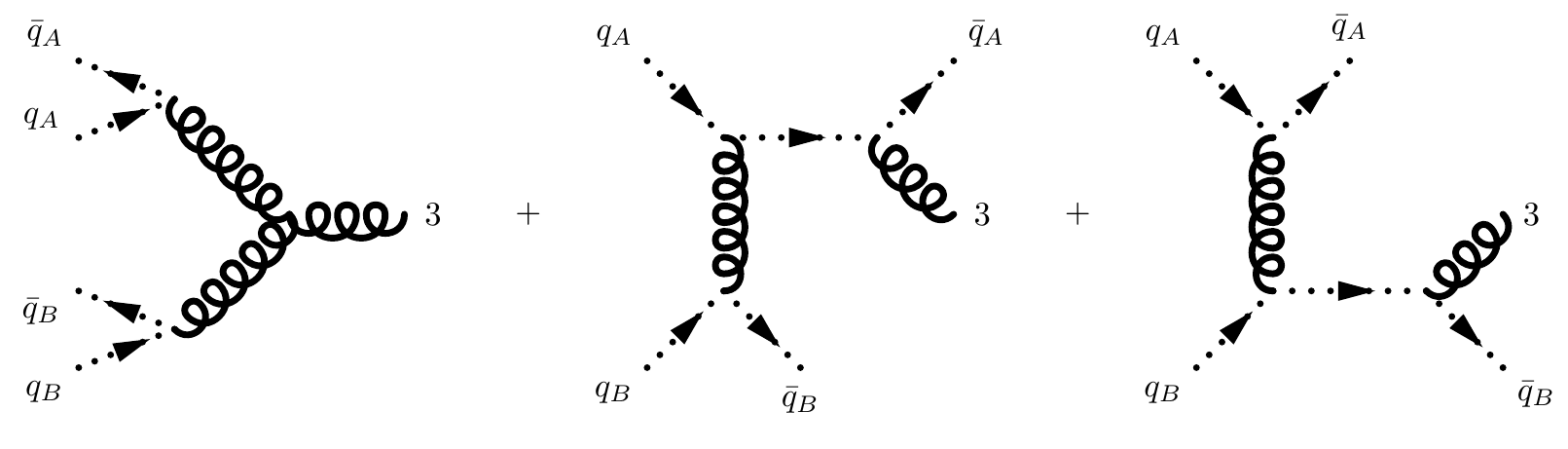}
\vspace{-0.5cm}
\caption{Feynman diagrams contributing to the amplitude used to define the generalized 3-gluon vertex.}
\label{fig:pggVertex}
\end{center}
\end{figure}

In order to calculate $\tilde{P}_{gg}$ splitting we needed to extend formalism
of~\cite{Catani:1994sq,Gituliar:2015agu} to the gluon case. This was achieved
in~\cite{Hentschinski:2017ayz} by generalizing definition of projector operators
and defining appropriate generalized 3-gluon vertex that is gauge invariant in
the presence of the off-shell momentum $k$. Definition of the generalized vertex
follows from application of the spinor helicity methods to the high-energy
factorization~\cite{vanHameren:2012uj,vanHameren:2012if,vanHameren:2013csa,vanHameren:2014iua,vanHameren:2015bba,vanHameren:2016bfc}
and can be obtained by summing the diagrams of Fig.~\ref{fig:pggVertex},
giving:
\begin{equation}
\label{eq:ggg_vertex} 
\Gamma^{\mu_1\mu_2\mu_3}_{g^*g^*g}(q,k,p') = \mathcal{V}^{\lambda
  \kappa \mu_3}(-q,k,-p') \, {d^{\mu_1}}_{\lambda} (q)\,
{d^{\mu_2}}_{\kappa}(k) +\, d^{\mu_1\mu_2}(k)\, \frac{q^2 n^{\mu_3}}{ n\cdot p'} -
d^{\mu_1\mu_2}(q)\, \frac{k^2 p^{\mu_3}}{ p\cdot p'} \, ,
\end{equation}
with $\mathcal{V}^{\lambda \kappa \mu_3}(-q,k,-p')$ being the ordinary 3-gluon QCD vertex,
and $d^{\mu_1\mu_2}(q)=-g^{\mu_1\mu_2}+\frac{q^{\mu_1}n^{\mu_2}+q^{\mu_2}n^{\mu_1}}{q\cdot n}$
is the numerator of the gluon propagator in the light-cone gauge.
More details on the exact procedure can be found in~\cite{Hentschinski:2017ayz}.

The definition of the projectors is more involved. We need to ensure that in the collinear
limit the new projectors reduce to the ones introduced by the CFP~\cite{Curci:1980uw}
and that the appropriate high-energy limit for the gluon splitting is also obtained.
A natural approach is to modify only the incoming projector $\mathbb{P}_{g,\,\text{in}}$, 
as the kinematics of the incoming momentum is more general, and keep the collinear outgoing
projector $\mathbb{P}_{g,\,\text{out}}$ unchanged (as the kinematics of the outgoing momentum is the same).
This is what was done by Catani and Hautmann~\cite{Catani:1994sq,Catani:1990eg} and later by
us~\cite{Gituliar:2015agu} when defining the projector for the calculation of the quark splitting
functions. In that case we used a transverse projector: $k_\perp^\mu k_\perp^\nu / k_\perp^2$
which, however, due to the more complicated structure of the 3-gluon vertex
$\Gamma^{\mu_1\mu_2\mu_3}_{g^*g^*g}$ can not be used any more. Instead we use, a more natural
(for high-energy factorization), longitudinal projector given by:
$\mathbb{P}_{g,\,\text{in}}^{\mu\nu} = y^2p^{\mu}p^{\nu}/k_{\perp}^2$. However, in order to satisfy
the requirements of $\mathbb{P}_{g}^s=\mathbb{P}_{g,\,\text{in}}\mathbb{P}_{g,\,\text{out}}$ being
a projector operator ($\mathbb{P}_{g}^s\otimes\mathbb{P}_{g}^s=\mathbb{P}_{g}^s$) we also need to modify
the outgoing projector. In the end we end up with the following gluon projectors:
\begin{equation}
\Pbb_{g,\,\text{in}}^{\mu\nu} =
-y^2\, \frac{p^{\mu} p^{\nu}}{k_\perp^2} \, , 
\quad\quad
\Pbb_{g,\,\text{out}}^{\mu\nu}  = 
-g^{\mu\nu} + \frac{k^\mu n^\nu + k^\nu n^\mu }{k\cdot n} - k^2\, \frac{n_\mu n_\nu}{(k\cdot n)^2} \, . 
\label{eq:HE_proj_new}
\end{equation}
It is easy to check that now $\mathbb{P}_{g}^s\otimes\mathbb{P}_{g}^s=\mathbb{P}_{g}^s$ holds.
Also the new outgoing projector is consistent with the collinear case and one can show that 
in the collinear limit the difference between $y^2\, p^\mu p^\nu/k_\perp^2$ and
$k^\mu_\perp k^\nu_\perp/k_\perp^2$ vanishes when contracted into the relevant vertices.
More details are provided in~\cite{Hentschinski:2017ayz}.

\section{Results}
Using the elements introduced in the previous section we can compute the transverse momentum
dependent gluon-to-gluon splitting function. We present here only the final result:
\begin{align}
  \label{eq:ggsplitting}
  \tilde{P}_{gg}^{(0)} (z, \qtt, \kt)
 &=
 2 C_A\, \bigg\{
 \frac{\qtt^4}{\left(\qtt-(1-z)\kt\right)^2[{\qtt^2+z(1-z)\kt^2}]} 
\bigg[\frac{z}{1-z} + \frac{1-z}{z}  +
\notag \\
& \hspace{-1.5cm}  + 
(3-4z) \frac{\qtt \cdot \kt}{ \qtt^2} + z(3-2z) \frac{\kt^2}{\qtt^2}
\bigg] + \frac{(1 + \epsilon)\qtt^2 z(1-z) [2 \qtt \cdot \kt + (2z -1)
  \kt^2]^2}{2 \kt^2 [\qtt^2+z(1-z)\kt^2]^2} \bigg\} \, ,
\end{align}
or after angular averaging (and setting $\epsilon=0$):
\begin{eqnarray}
\bar{P}_{gg}^{(0)}\left(z, \frac{\kt^2}{\qtt^2} \right) 
&=&
C_A\, 
\frac{\qtt^2}{\qtt^2+z(1-z)\kt^2}
\bigg[
\frac{(2-z)\qtt^2+(z^3-4z^2+3z)\kt^2}{z(1-z)\left|\qtt^2-(1-z)^2\kt^2\right|}
\notag \\
&+& 
   \frac{(2z^3-4z^2+6z-3)\qtt^2+z(4z^4-12z^3+9z^2+z-2)\kt^2}{(1-z)(\qtt^2+z(1-z)\kt^2)}
\bigg] \, .
\end{eqnarray}

Now we can explicitly check the corresponding kinematic limits. 
In the collinear case this is straightforward, since the transverse integral in
Eq.~\eqref{eq:TMDkernelDefINI} is specially adapted for this limit. 
In particular, one easily obtains the real part of the DGLAP gluon-to-gluon splitting 
function:
\begin{align}
\label{eq:7}
\lim_{\kt^2 \to 0} \bar{P}_{ij}^{(0)}\left(z, \frac{\kt^2}{\qtt^2}\right) &= 2\, C_A\, \left[ \frac{z}{1-z} + \frac{1-z}{z} + z\,\left(1-z\right) \right] \, .
\end{align}
In order to study the high-energy and soft limit it is convenient to change to the following variables:
$\ptt = \frac{\kt - \qt}{1-z} = \kt - \frac{\qtt}{1-z}$, then in the high-energy limit ($z\to0$)
we obtain:
\begin{align}
  \label{eq:8}
  \lim_{z\to 0} \hat K_{gg} \left(z, \frac{\kt^2}{\mu^2}, \epsilon, \alpha_s \right)  
&=
\frac{\alpha_s C_A}{\pi (e^{\gamma_E}\mu^2)^\epsilon}\int \frac{d^{2 + 2 \epsilon} \ptt}{\pi^{1 + \epsilon}} \Theta\left(\mu_F^2 - (\kt - \ptt)^2\right) \frac{1}{\ptt^2} \notag
\end{align}
where the term under the integral is easily identified as the real
part of the LO BFKL kernel.
Additionally, one can check that in the angular ordered region of phase space,
where $\ptt^2 \to 0$, we reproduce the
real/unresummed part of the CCFM kernel~\cite{Ciafaloni:1987ur,Catani:1989yc,CCFMd}:
$\frac{1}{z} + \frac{1}{1-z} + \mathcal{O}\left(\frac{\ptt^2}{\kt^2}\right)$.

\section{Summary}
The main result of this paper is the calculation of a transverse
momentum dependent gluon-to-gluon splitting function.  The splitting
function reduces both to the conventional gluon-to-gluon DGLAP
splitting in the collinear limit as well as to the LO BFKL
kernel in the low $x$/high energy limit; moreover the CCFM
gluon-to-gluon splitting function is re-obtained in the limit where
the transverse momentum of the emitted gluon vanishes, {\it i.e.} if
the emitted gluon is soft.  The derivation of this result is based on
the Curci-Furmanski-Petronzio formalism for the calculation of DGLAP
splitting functions in axial gauges.  To address gauge invariance in
the presence of off-shell partons, high energy factorization adapted
for axial gauges has been used to derive an effective production
vertex which then could be shown to satisfy current conservation.
The next step in completing the calculation of TMD splitting functions
is the determination of the still missing virtual corrections.
With the complete set of splitting functions at hand, it will be finally
possible to formulate an evolution equation for the unintegrated (TMD)
parton distribution functions including both gluons and quarks. 

%

\bibliographystyle{utphys}
\bibliography{references}

\end{document}